\documentclass[twocolumn,showpacs,showkeys,nobibnotes,preprintnumbers,prl,aps,apssymb]{revtex4-1}
\usepackage{amsmath,amssymb}
\usepackage{graphicx}
\usepackage{dcolumn}
\usepackage{bm}
\usepackage{psfrag}
\usepackage{subfigure}
\newcommand{\ber}{\begin{eqnarray}}
\newcommand{\eer}{\end{eqnarray}}
\newcommand{\bea}{\begin{equation}}
\newcommand{\eea}{\end{equation}}

\begin{document}

\title{Directed transport in equilibrium}
\author{A. Bhattacharyay}
\email{a.bhattacharyay@iiserpune.ac.in}
\affiliation{Indian Institute of Science Education and Research, Pune, India}
\thanks{Ph.no. +91 (0)20 25908095}
\date{\today}

\begin{abstract}
\begin{center}
{\bf abstract}                 
\end{center}
We investigate a symmetry broken dimer constrained to move in a particular direction when in contact with a heat-bath at a constant temperature. The dimer is not driven by any external force. The system gains kinetic energy from the heat-bath. The symmetry broken system can use this energy in directed transport. At the hard core collision limit between the particles of the dimer, we show by exact analytic calculations and complementary numerical results that the dimer undergoes steady directed transport. Our observation, being consistent with the {\it second law of thermodynamics}, {\it detailed balance} etc leads to new physical understanding to which much attention has not been paid.
\end{abstract}
\maketitle

\pagebreak

{\bf Introduction}\\
The Brownian Ratchet (BR) paradigm provides foundation to the understanding of symmetry-broken driven directed transport in contact with a heat-bath \cite{ast,rei,han,han1}. BR paradigm is based on the concepts of Smoluchowski-Feynman ratchet that illustrates a typical case where extracting energy from the random motions of a heat-bath alone is impossible due to violation of the second law of thermodynamics \cite{feyn}. In general, BR paradigm claims that thermal noise assists driven transport in the presence of a repeated spatial asymmetry. Attempts have been made in understanding biological transports in the light of BR paradigm by invoking numerous variants of the basic BR model in order to match the efficiency of the actual biological systems \cite{vale,ast1,oka,gei,cil,von,mogi,fog,kum}. However, the efficiency matching is quite poor till date. There are many other areas, like flow through micro-channels, nanopores, optical lattices etc., where the BR paradigm finds its application to the noise assisted driven transport phenomena \cite{han1,jone,pale,sjol}. There also exists a variant of conventional BR, called an Intrinsic Ratchet (IR) model, where the internal degree of freedom is not coupled to an external global symmetry breaking field. Interestingly, such IR systems with microscopic broken symmetry are capable of showing macroscopic directed transport in the presence of a drive that disturbs the equilibrium of the system intermittently \cite{bro,van,nor}.
\par
It is generally believed that, the Smoluchowski-Feynman ratchet model denies any filtering of random motions of a heat bath under equilibrium conditions. Moreover, when the dynamics of an interacting system of particles is cast in the (over-damped) Langevin equation (LE) form \cite{lang,blan,risk}, the position coordinate becomes the relevant variable. The demand of a stationary probability distribution of the variable of this general LE, as a condition for the equilibrium of the system with the heat-bath, would (arguably) preclude any directed transport of the system. Because, any average directional drift of the system might mean a drift in the stationary probability distribution of the position coordinates which will be in conflict with the demand of stationary distribution of position variables. But, as one can reasonably guess, (i) the second law of thermodynamics would not forbid such a constant drift so long as no energy is being extracted from the system, (ii) the detailed balance on a uniformly moving frame is maintained due to homogeneity of space which makes interactions between particles functions of their distance of separation and (iii) nonexistence of time reversal invariance for a moving system is not inconsistent with second law of thermodynamics (Loschmidt's paradox). To our knowledge, possibility of such directed transport of a mechanical system in equilibrium with a heat-bath (needless to say in the absence of any external driving) has not been systematically explored so far. In the present paper, we investigate one generic model for directed transport in equilibrium with a heat bath.
\par
We consider a microscopic system with internally broken symmetry like the IR, but, unlike IR its not driven for directed transport. Rather, we investigate a set of completely different issues to understand what happens when such a symmetry-broken microscopic object is brought in contact with a heat-bath and the locally broken symmetry is not allowed to relax. If the origin of the broken internal symmetry of the system is consistent with conditions of equilibrium with the heat-bath, the system would definitely have a preferred direction of motion and must also equilibrate at large times. If it does not equilibrate, that is a violation of second law of thermodynamics in itself, because, under such ever-persisting non-equilibrium situation one may extract energy from a heat-bath for ever without external driving. In the following, we first identify a general situation in which such a system would show directed transport. We show exact analytic results for elastic hardcore collisions and provide numerical support to it. We identify the damping or dissipation as the key physical ingredient required for symmetry breaking directed motion in equilibrium. We conclude with a discussion of our observations in relation with second law, detailed balance, entropy production etc.
\\ 
{\bf Model}\\
The dynamics for our model system is governed by the set of equations
\ber\nonumber
\frac{\partial^2x_1}{\partial t^2}=&-&(1-\beta)\frac{\partial x_1}{\partial t}-\alpha(x_1-x_2)\\ &+&\sqrt{2(1-\beta)k_BT}\eta_1(t)-\frac{\partial \Phi (x_1-x_2)}{\partial x_1}\\\nonumber
\frac{\partial^2x_2}{\partial t^2}=&-&\frac{\partial x_2}{\partial t}+\alpha(x_1-x_2)+\sqrt{2k_BT}\eta_2(t)\\ &-&\frac{\partial \Phi (x_1-x_2)}{\partial x_2}.
\eer 
The mass of the particles at $x_1$ and $x_2$ ($x_1>x_2$) are unity and damping constants are $(1-\beta)$ and 1 respectively. The damping can be different on different particles depending upon size, shape etc. of the particles, where by particle we do not mean dimensionless point objects. The random forces on these particles $\eta_1$ and $\eta_2$ are Gaussian white noise with the moments $\left <\eta_i(t)\right >=0$ and $\left <\eta_i(t_1)\eta_j(t_2)\right >=\delta_{ij}\delta(t_1-t_2)$. The particles are bound by a spring force of force constant $\alpha$. In the last term $\Phi(x_1-x_2)$ is the collision potential taking care of the excluded volume interaction keeping $x_1>x_2$ for ever. The strength of the thermal noises are explicitly mentioned in accordance with the fluctuation-dissipation relation so that the particles see the same temperature $T$ of the heat-bath where $k_B$ is the Boltzmann constant. Similar models with broken symmetry due to different damping have been considered in ref.\cite{geh,ari}, but, systems are driven there.
\par
Let us think of a situation where the collision potential $\Phi(x_1-x_2)$ is so steep that the actual collision time (the time in which particles are in contact with each other) is very very small compared to the slow time scale over which the over-damped dynamics of the system takes place. This situation allows us to separate the dynamics into two parts. The inertial part only takes care of the collision forces and the over-damped part takes care of the rest. The inertial part on the center of mass (CM) $X=(x_1+x_2)/2$ and internal coordinate $Z=x_1-x_2$ would look like
\ber
\frac{1}{2}\ddot{Z} &=& -\frac{\partial \Phi (Z)}{\partial Z}\\
\ddot{X} &=& 0
\eer
The inertial part does not contain other forces because of the following reasons. The damping is particularly defined at a larger time scale and there is every reason to take the stochastic force averaging out over several such intermittent collisions because they are not correlated. In the following where we would be introducing the corrections occurring from the collisions to the average velocity of the CM and $Z$, we would actually be introducing the average contributions over several such collisions. Moreover, this is a plausible hypothesis because we can actually keep  $\alpha$, in the present model, arbitrarily small which would in turn make average time gap between collisions large and that can actually make the stochastic contribution at collisions uncorrelated. So, based on the above logic of considering time scale corresponding to the harmonic interaction large Eq.4 clearly indicates that there will be no contribution to the CM velocity from the collision whereas Eq.3 indicates a contribution to the relative velocity of the particles coming from the collision which would be just in the opposite direction to that coming from the harmonic part. For the system to remain in equilibrium the $<\dot{Z}>$ must vanish and that would happen due to cancellation of contributions coming to it from the harmonic interaction and this collision part. In what follows we would take care of this short time scale dynamics through the boundary conditions imposed to the large time scale dynamics. Its important to note that Eq.1 and 2 with all the forces present simultaneously does not represent our model. Our model particularly relies on the non-simultaneous acting of the collision and damping like slow forces. This separation of dynamics is crucial to get the expected results because it will be shown in the following that the effective symmetry breaking of the slow harmonic interaction by the damping keeping the collision symmetry intact is the basis of getting the directed motion of the model. It is also important to note, at this stage, that this somewhat idealized model we are going to analyze is entirely because of the simplicity. Based on this model, we would understand a more realistic situation taking into account inelastic collisions of the particles where we can actually have such conditions of idealized hardcore collisions relaxed.
\par 
The equilibrium of different particles (in shape/size etc.) with the same heat-bath ensures broken symmetry and this is a key aspect of the construction of the present model. Considering the over-damped (slow) part of the dynamics to cast the model into a standard coupled Langevin equation \cite{lang,blan} and moving on to the internal coordinate and CM, we rewrite the above set of equations in a more convenient form as
\ber
\dot{Z}=-\frac{(2-\beta)}{1-\beta} \frac{\partial U(Z)}{\partial Z}+ \xi_Z(t)\\
\dot{X}=-\frac{\beta}{2(1-\beta)} \frac{\partial U(Z)}{\partial Z} +\xi_X(t),
\eer 
where $U(Z)=\frac{\alpha}{2} Z^2$. The modified random forces $\xi$s are functions of $\eta$s and are of zero average. Explicitly, expressions of the noises are $\xi_Z=\sqrt{\frac{2k_BT}{1-\beta}}\eta_1(t)-\sqrt{2k_BT}\eta_2(t)$ and $\xi_X=(\sqrt{\frac{2k_BT}{1-\beta}}\eta_1(t)+\sqrt{2k_BT}\eta_2(t))/2$. The second moments of the random forces are given as $\left <\xi_Z(t_1)\xi_Z(t_2)\right >=2Tk_B\frac{2-\beta}{1-\beta}\delta(t_1-t_2)$, $\left <\xi_X(t_1)\xi_X(t_2)\right >=Tk_B\frac{2-\beta}{2(1-\beta)}\delta(t_1-t_2)$ and $\left <\xi_Z(t_1)\xi_X(t_2)\right >=-\beta k_B T\delta(t_1-t_2)$. The steady state probability distribution for the $Z$, when $U(Z)$ is well behaved, can be readily given as $P(Z)=N\exp{(-U(Z)/k_B T)}$ (N is normalization constant) with a sharp fall to zero value at $Z=Z_l$ which is the boundary condition imposed in view of the steep excluded volume interaction (fast collision process). This probability distribution can be used in general to write down an expression for the average velocity of the CM as
\ber\nonumber
V=<\dot{X}> &=& -\frac{\beta}{2(1-\beta)}<\frac{\partial U(Z)}{\partial Z}> \\ &=&\frac{\beta k_B T}{2(1-\beta)}\int_{Z_l}^\infty{\frac{dP(Z)}{dZ}dZ}.
\eer
Important to note that, in the above integral we would not take into account the delta function contribution that comes near $Z=Z_l$. This delta function contribution represents the contribution of the collision potential to the velocity of CM which we know from Eq.4 to be identically equal to zero. The only contribution that would come to the velocity of the CM is from the harmonic part of the interaction. Remember that, its the interactions over the span of the harmonic potential only which contributes to the development of the velocity of the CM and during the collisions between the particles constituting the dimer that velocity remains conserved. But, for $<\dot{Z}>$ we must take into account the delta function contribution to the integral at $Z=Z_l$ which would make $<\dot{Z}>=0$ although $<\dot{X}>\neq 0$. We would take into account the delta function contribution for the average $\dot{Z}$ because this part representing the collision indeed contributes to the relative velocity of the particles of the dimer with respect to each other. Its interesting to note that, consideration of a very steep collision potential amounts to abruptly bringing the $P(Z)$ from its maximum value to zero over a very short range of $Z$ near $Z=Z_l$ which is at the origin of consideration of the delta function contribution. Also note that, since only the harmonic interaction is contributing to the velocity of the CM, $<\dot{X}> $ is proportional to $ <Z>$. The reason the harmonic interaction is contributing to the average CM velocity is that its effectively asymmetric and the symmetries of the collision interaction prevents it from having any contribution to the CM velocity.
\\
{\bf Results}\\
\begin{figure}
\subfigure []
{\includegraphics[width=4.5 cm,angle=-90]{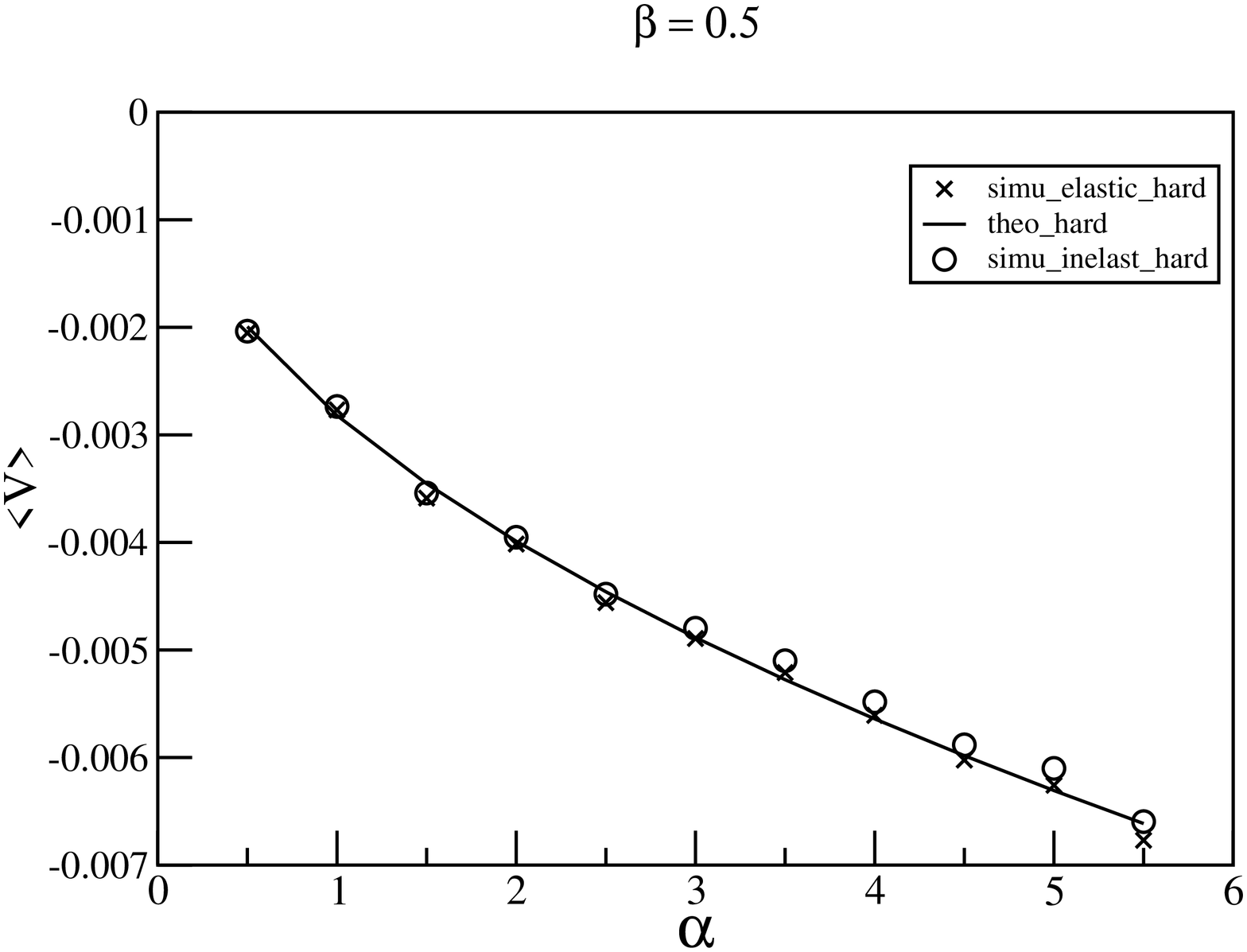}
\label{Fig:a}}
\subfigure []
{\includegraphics[width=4.5 cm,angle=-90]{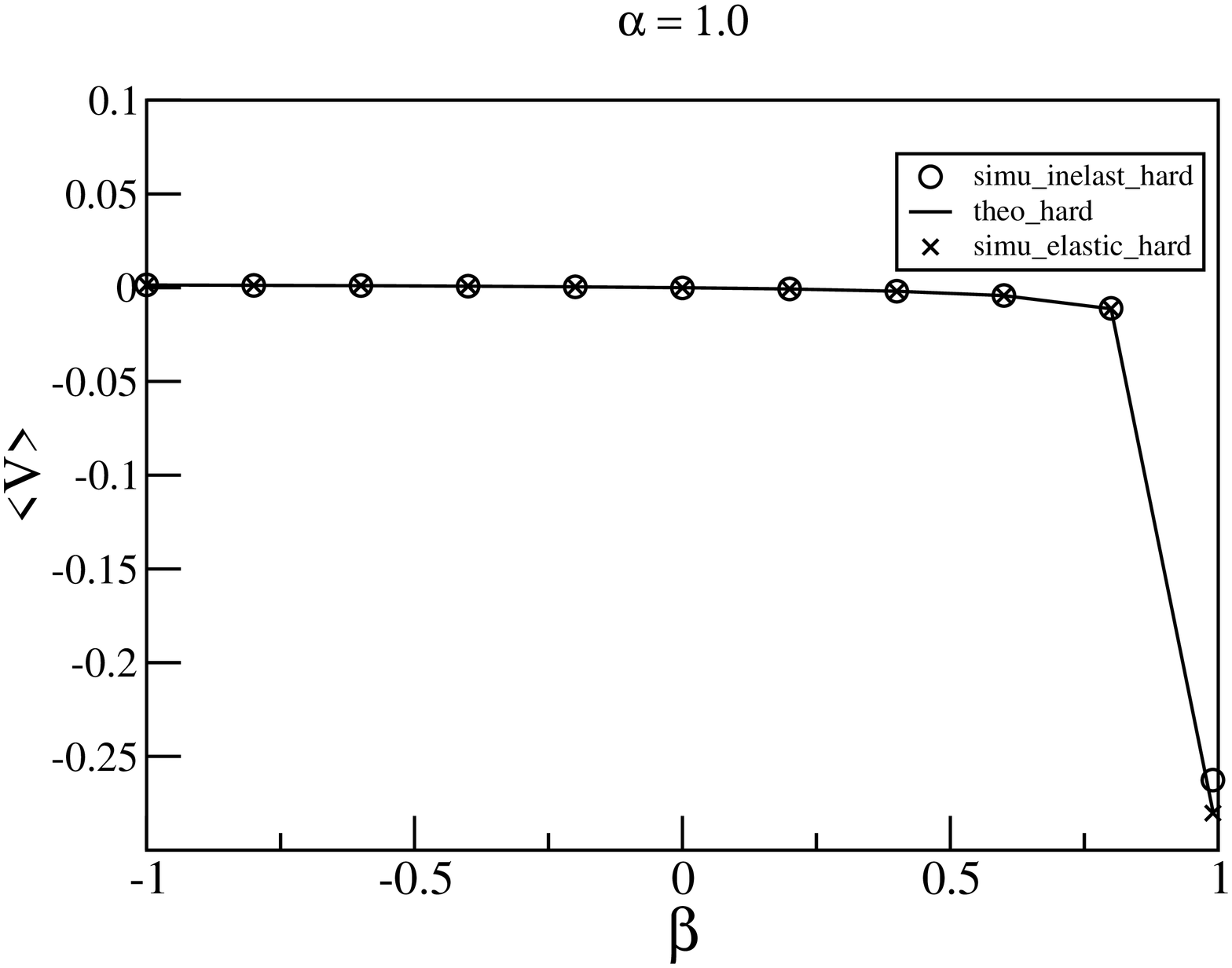}
\label{Fig:b}}
\caption [Figure 1]{{\bf Average velocity V of CM plotted against force constant {\boldmath $\alpha$} and damping constant {\boldmath $\beta$}}: (a) Average velocity vs $\alpha$ at $\beta=0.5 $ and (b) Average velocity vs $\beta$ at $\alpha=1.0. $}
\end{figure}
From the over-damped dynamics one obtains an equilibrium distribution of $Z$ as $P(Z)=Ne^{\frac{-\alpha Z^2}{2k_B T}}$ within the range $0<Z<\infty$ (here $Z_l=0$ and at this point $P(Z)=0$ is the boundary condition). If the system indeed attains this steady state distribution at large times from any initial conditions, the steady average velocity of the CM will be given by 
\bea
V=-\frac{\beta}{1-\beta}\sqrt{\frac{k_B T \alpha}{2\pi}},
\eea
which is an exact result.
\par
A direct numerical simulation of Eq.5 and 6 with this elastic hardcore collision is actually possible and reveals numbers in excellent agreement with the exact analytic results. In Fig.1(a), we have plotted the average velocity of CM against the parameter $\alpha$ and in Fig.1(b) we have plotted the same against $\beta$. While doing the numerical simulation, we have implemented the repulsion between particles with the following rules. Since, for particles of same mass (as has been considered here), there is an exchange of velocity at elastic hard-core collisions, we simply exchange the position of the particles following any overshooting across each other during the simulation. This simple implementation of boundary conditions, while sequentially updating particle positions, takes care of the perfect hard-core collision preserving the momentum at collisions. Similar agreements have also been seen at the perfect inelastic collision limit where we put both the particles at the CM following any overshooting and allow them to move in their respective sides only in the next time step if no overshooting is happening again. The perfect inelastic collision also showing similar results due to the fact that it also does not mess up with the CM momentum conservation at collisions. The case of this inelastic collision is the key to understand the physically realizable situation for such motions. We will discuss about it at length in the following, but, here we would like to mention that under the scope of our present model, where the temperature of the bath is kept constant despite implementing inelastic collision between particles in the simulation, we are considering that the particles are relaxing back quickly by giving back the energy to bath following an inelastic collision. The excellent agreement between the theory and the simulation indicates that the system has indeed attained its equilibrium distribution in $Z$ which clearly comes out to be the case in Fig.2 where numerical and analytical distributions of $Z$ are compared. We have checked this comparison of theory and simulation for a range of $\alpha$ and $\beta$ and those are quite consistent. The excellent matching of probability distributions indicates that corresponding Fokker-Planck equation (FPE) works fine for a $U(Z)$ having discontinuity at the boundary. In Fig.3 we have plotted few trajectories of the moving dimer for the illustration of the average directed motion of the system at various parameter values. In Fig.5 we have shown a comparison of $<Z>$ against $\alpha$ as obtained from theory and numerics. In all these simulations the $k_BT=0.00005$, the time step is $\bigtriangleup t = 0.0001$, a single run is over $10^7$ such time steps and the averaging is done over 10 such runs starting from the same initial conditions for each set of parameters. Note that, the characteristic time scale for this system being $\beta/\alpha$ is of the order of $0.01$.
\begin{figure}
{\includegraphics[width=5 cm,angle=-90]{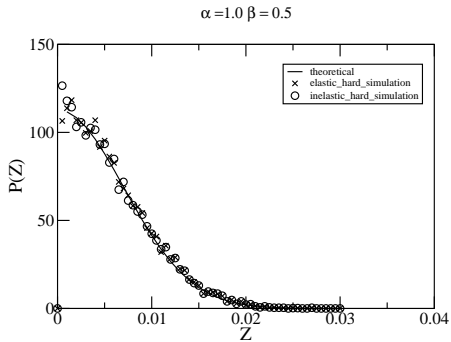}}
\caption [Figure 2]{Comparison of numerical and analytical probability distribution of $Z$: Distribution of $Z$ for $\alpha=1.0$ at $\beta=0.5 $. }
\end{figure}
\begin{figure}
{\includegraphics[width=5 cm,angle=-90]{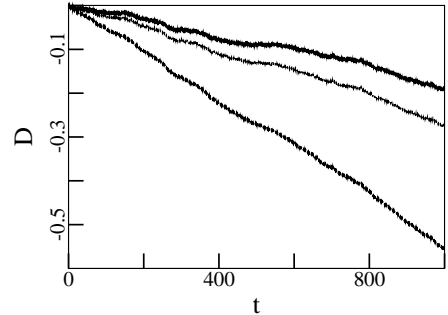}}
\caption [Figure 3]{ Trajectories of the dimer for hard-core collisions: $\beta=0.5 $ and $\alpha = 0.5$, $1.0$ and $4.0$ from top to bottom.}
\end{figure}
\par
Consider the fact that, the dimer is maintaining a uniform velocity in the presence of damping. So, what balances this damping to help the system maintain its velocity? Eq.6 depicts the over-damped dynamics of the CM in the presence of two forcing terms - the one originating from the harmonic interaction between the particles and the other one being the random force due to the bath degrees of freedom and that is of zero average. Obviously, there is no contribution to the average velocity of the CM from the random force part because it has a zero average. Also remember that, in the case of hardcore collisions the contribution from the collision forces to the average CM velocity is zero (as we have already shown). To understand the effective force working against the damping refer to Fig.4a where we have plotted the effective spring potentials as seen by the particles at $x_1$ (right) and $x_2$ (left). To make the picture more revealing we have plotted the potentials in such a way that the inward direction of the resulting forces are evident. The effective spring constant for the particle at $x_1$ being $(1-\beta)^{-1}$ times bigger than that of the particle at $x_2$, there is an effective force on the composite system from $x_1$ towards $x_2$. This effective force would on average continuously balance the damping of the CM velocity. Interesting to note that, the velocity picked up by the system is in the direction of this effective force which must have happened at the initial transients when the average uniform velocity was being reached for which the damping (which is proportional to velocity) on average is the same as the effective force.
\par
Refer to the schematic diagram of Fig.4b to understand why there is no velocity of the CM when there is a soft-core conservative collision (collision time is of the order of the characteristic time scale of over-damped dynamics) between the particles. Here, we have considered a collision potential of the simple form $\Phi(Z)=(a-bZ)$ for $0\leq Z\leq \epsilon$ where $\epsilon$ is the width of the collision potential. $a$ and $b$ are suitably large numbers (depending on $k_BT$) to make sure there is no passage of particles through each other. Moreover, we consider that the damping due to the bath works equally when particles actually collide with each other as when they are on flight. In such a situation, since, the attractive (harmonic) and repulsive (collision) effective forces between particles are similarly renormalized by their respective damping constants, individually on average there will be a force balance for each of the particles. Its not difficult to understand how the gain of the CM velocity in one direction is counterbalanced in such a situation when the collision potential is conservative. The velocity of approach and the velocity of the separation of the particles just at the point of colliding would be the same and the velocity of separation would now counterbalance the gain in CM velocity that happened during the approach till the next approach is initiated as the particles are gone far apart. Interesting to note that, in this case of soft conservative collisions in the presence of the symmetry breaking damping, the velocity of separation is achieved by the velocity reversal and not by the velocity exchange. As a result the velocity of separation would contribute as much to the velocity of CM as the velocity of approach does, but, it would do that in just the opposite direction. At this point, consider what happens if the collision is perfectly inelastic. The velocity of separation being zero there is nothing to counterbalance the CM velocity gained by the help of velocity of approach. So, under the conditions that the energy absorbed by the particles at inelastic collision are quickly enough given back to the bath and no energy is lost from the composite system of the bath and the dimer the system would show directed transport even for a soft-collision between the particles. Indeed, we have seen the hard-core perfectly inelastic collision to produce exactly the same results as the hardcore elastic collision in our numerical simulation. Let us discuss some subtleties of this inelastic collision case in the following paragraph a little more elaborately.

\begin{figure}
{\includegraphics[width=5 cm,angle=-90]{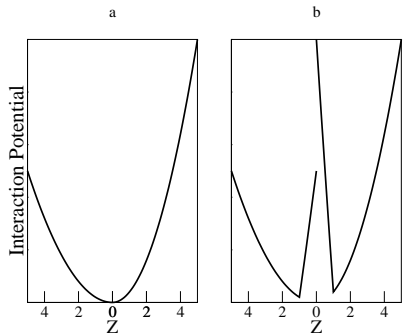}}
\caption [Figure 4]{ Schematic diagram of effective potentials.}
\end{figure}

\begin{figure}
{\includegraphics[width=5 cm,angle=-90]{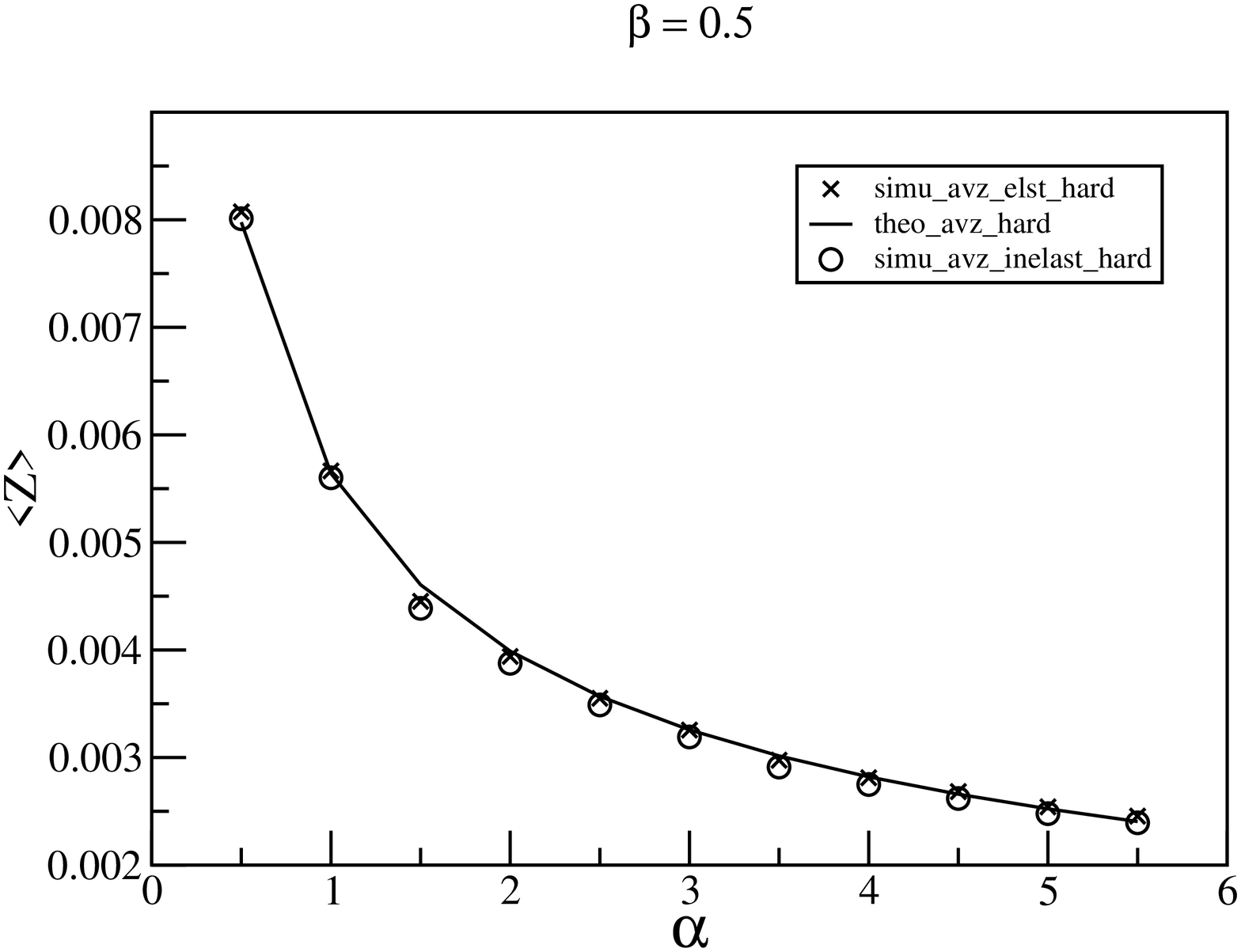}}
\caption [Figure 4]{ Plot of $<Z>$ against $\alpha$.}
\end{figure}

\par
Note that, for soft collision potentials there may come an explicit contribution of the collision potential in Eq.5 and 6 because now the resulting collision forces can be a part of the over-damped dynamics. If we think in terms of the delta function contribution to the integral in Eq.7, then, those are now there for the velocity of the CM and that of the $Z$ making both vanish. At the origin of the collision contribution to the average velocity of the CM is the effective symmetry breaking of the collision force by the unequal damping exactly the same way as that happened for the harmonic interaction. There is an way out though, and that involves the possibility of having a different damping during collisions (when particles are in touch with each other). Since the environment when the particles are at collision is different from that when they are in flight the damping can actually be different at these two times. Furthermore, there can be inelastic collisions which is absolutely of a different origin than the fluid friction and which can make the overall damping the system of particles experience during collisions become different from that when they are in flight. One has to keep in mind that if there is inelastic collision the energy ultimately has to be given up by the system to the bath and the system cannot behave as an infinite heat sink. Thus, the system has to equilibrate with the bath at large times but this equilibration time should now be different depending upon the relaxation time of the internal vibration modes of the particles etc. Most important to note that, under such conditions, the fluctuation dissipation must hold good when the system is in equilibrium. But, more plausibly, it would hold good at a larger time scale with an average damping which is an average over the different damping the particles are actually undergoing while on flight and at collision. This is an equilibrium situation where there are more long lived fluctuations involved then usual due to the role played by other internal degrees of freedom of the system equilibrating with the bath. This kind of a situation, to my knowledge, is not quite well studied so far. This is a situation which again comprises of two time scales within the average time scale over which the FDT is defined and one can see a variation of the damping at these two time scales. Having two damping, it opens up the possibility, in a similar way as discussed above, of having persistent broken symmetry in the system in equilibrium and, hence, directed motion for soft collisions as well. Probably this is the scenario more easier to probe experimentally if one can ensure the isolation of the system and the bath from the surroundings.
\\
{\bf Numerical result with inertial term added}\\
In the following we are going to present some results obtained from the direct numerical simulation of the full model containing inertial terms and we will see that, the basic observation of the over-damped (simple) model is also present there. We will see that, the velocity of the CM scales as squire root of the force constant $\alpha$ and $k_BT$ and picks up the write direction as is evident from the discussion of the symmetry breaking. The absolute value of the velocity is somewhat smaller than that obtained from the over-damped limit for obvious reasons of having velocity fluctuations considered in the full inertial model in the presence of the masses. However, an analytical solution of the full model with inertial terms, till now, has not been possible to do because of the strong nonlinearity introduced by the collision potentials. The model is the following
\ber\nonumber
\frac{\partial x_1}{\partial t} &=& v_1\\\nonumber
\frac{\partial v_1}{\partial t} &=& -(1-\beta)v_1-\alpha(x_1-x_2) + \sqrt{2(1-\beta)k_BT}\eta_1\\\nonumber
\frac{\partial x_2}{\partial t} &=& v_2\\\nonumber
\frac{\partial v_2}{\partial t} &=& -v_2+\alpha(x_1-x_2)+\sqrt{2k_BT}\eta_2.
\eer   
While simulating the above model, we will exchange the velocity and position of particles at each overshooting considering them to be of the same mass. This is the implementation of the hard collision happening in the absence of the damping as has been explained in details in the context of the over-damped model. For the full model, we are also exchanging velocity on top of doing the same for the position. This is a very simple way of implementing the basic requirement of our prototypical model of having non-simultaneous working of the damping (slow) and the collision (hard) interactions in the numerical simulation which actually proves technically difficult in the analytical handling of the full inertial model. 

\begin{figure}
{\includegraphics[width=5 cm,angle=-90]{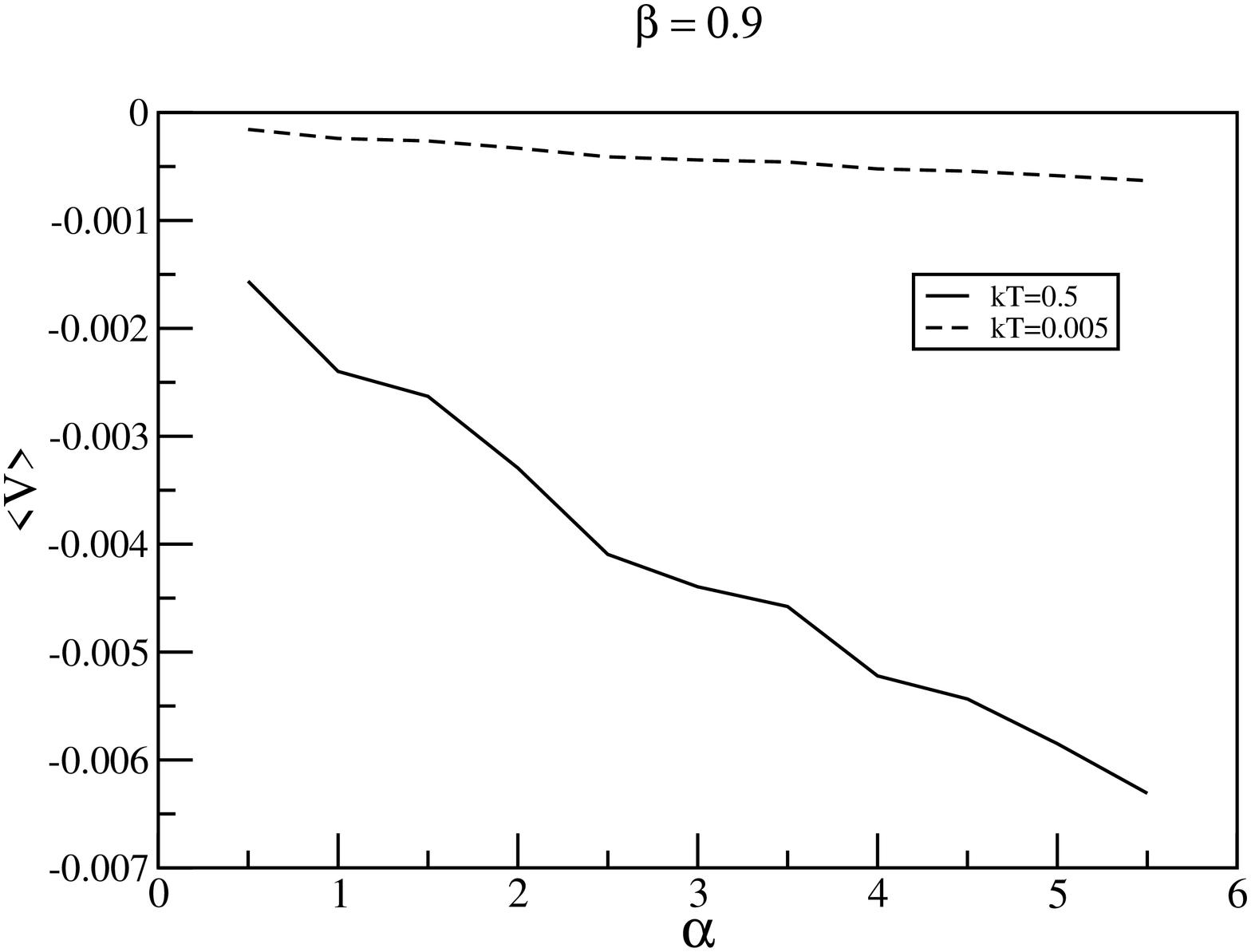}}
\caption [Figure 4]{ Plot of $<V>$ vs $\alpha$.}
\end{figure}

\begin{figure}
{\includegraphics[width=5 cm,angle=-90]{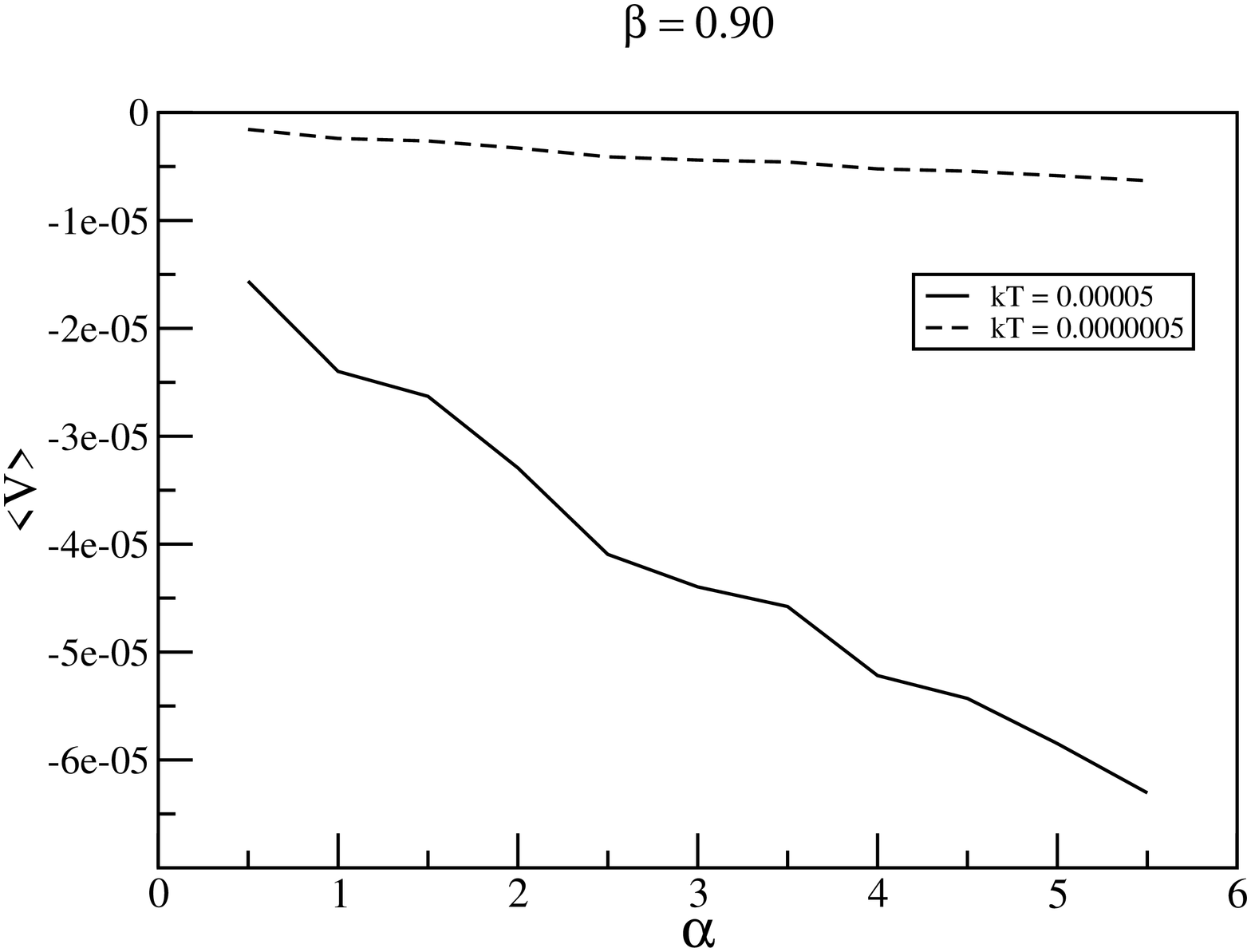}}
\caption [Figure 4]{ Plot of $<V>$ against $\alpha$.}
\end{figure}

\par
In Fig.6a and 6b we have plotted the average velocity of the CM against $\alpha$. Fig.6a shows plots for the value of $k_BT = 0.5$ (lower one) and $k_BT=0.005$ (upper one), whereas, Fig.6b shows plots of the same for $k_BT=0.00005$ (lower) and $k_BT=0.0000005$ (upper). Its clear from these figures that the system is showing similar results as the over-damped system but the absolute value of the CM velocity is generally smaller by 2 orders of magnitude than the over-damped one. The masses taken here are unity as is obvious from the model mentioned above. The simulations have been done with a time step $\bigtriangleup t = 0.01$ for $10^8$ steps for every set of parameters and then averaging over $50$ such runs each starting from the same initial conditions. If we take the magnitude of the velocity and plot the graphs on log scales against $\alpha$ (Fig.7) and $k_BT$ (Fig.8) we immediately see that the velocity scales as $\sqrt{\alpha}$ and $\sqrt{k_BT}$ as is predicted by the over-damped model which confirms the validity of our exact over-damped analysis handling two different time scales. Actually, the exponent of $\alpha$ varies between 0.5 and 0.6 but we think a better averaging would fix it close to 0.5. The scaling with $\alpha$ remaining unchanged is important in view of the fact that its the harmonic interaction which basically drives the system. The scaling with $\sqrt{k_BT}$ is important in view of the fact that the symmetry broken 1D system being bound to move in a single direction should also respect equipartition in equilibrium which is a symmetry of the equilibrium itself. The deviation in the amplitude of the velocity from the over-damped expression cannot be fixed without being able to do the analytic work on this model. Anyway, having a basic confirmation of important scalings from the numerics of the full model lends support to our simplified analysis with the over-damed model.

\begin{figure}
{\includegraphics[width=5 cm,angle=-90]{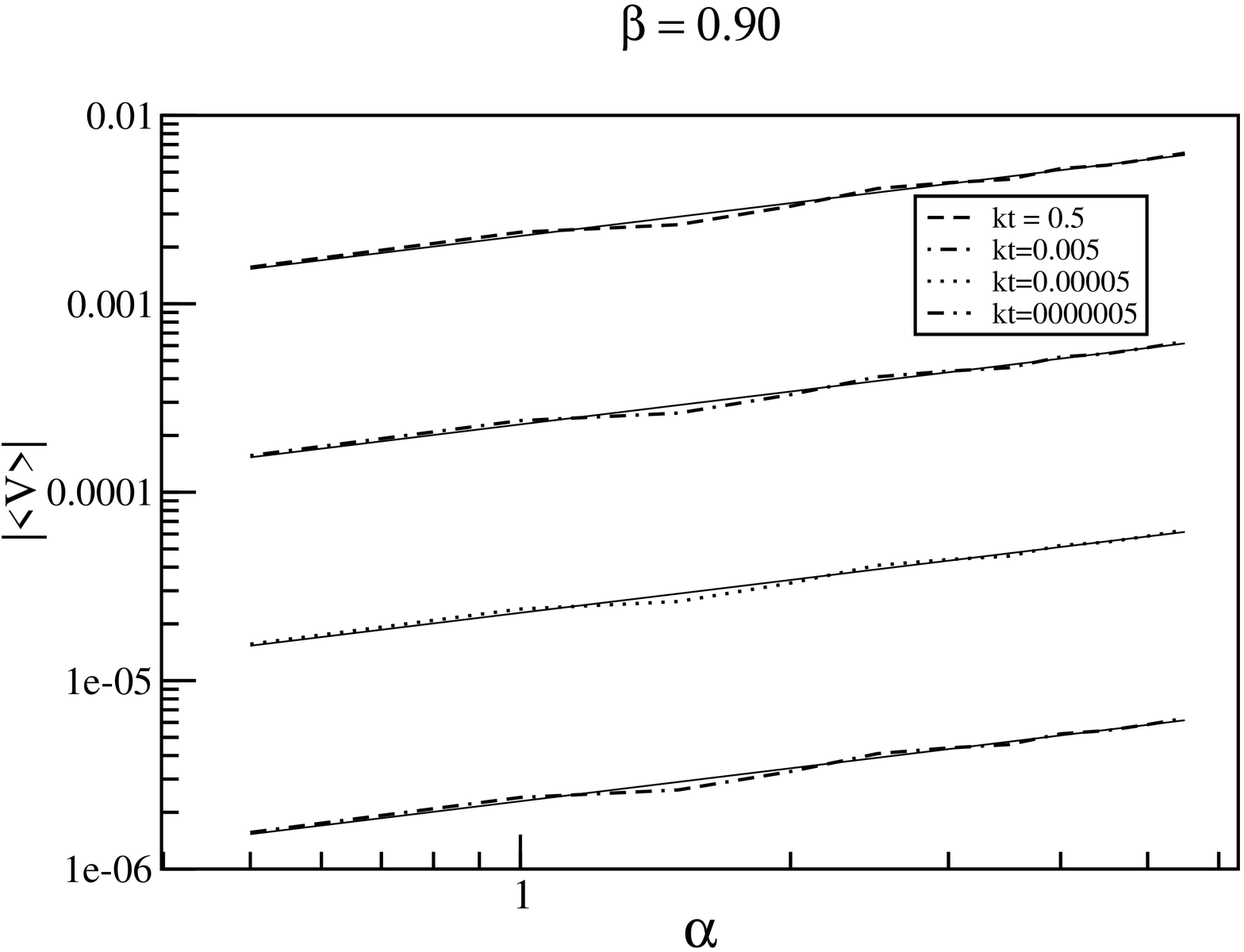}}
\caption [Figure 4]{ Scaling of $|<V>|$ with $\alpha$.}
\end{figure}

\begin{figure}
{\includegraphics[width=5 cm,angle=-90]{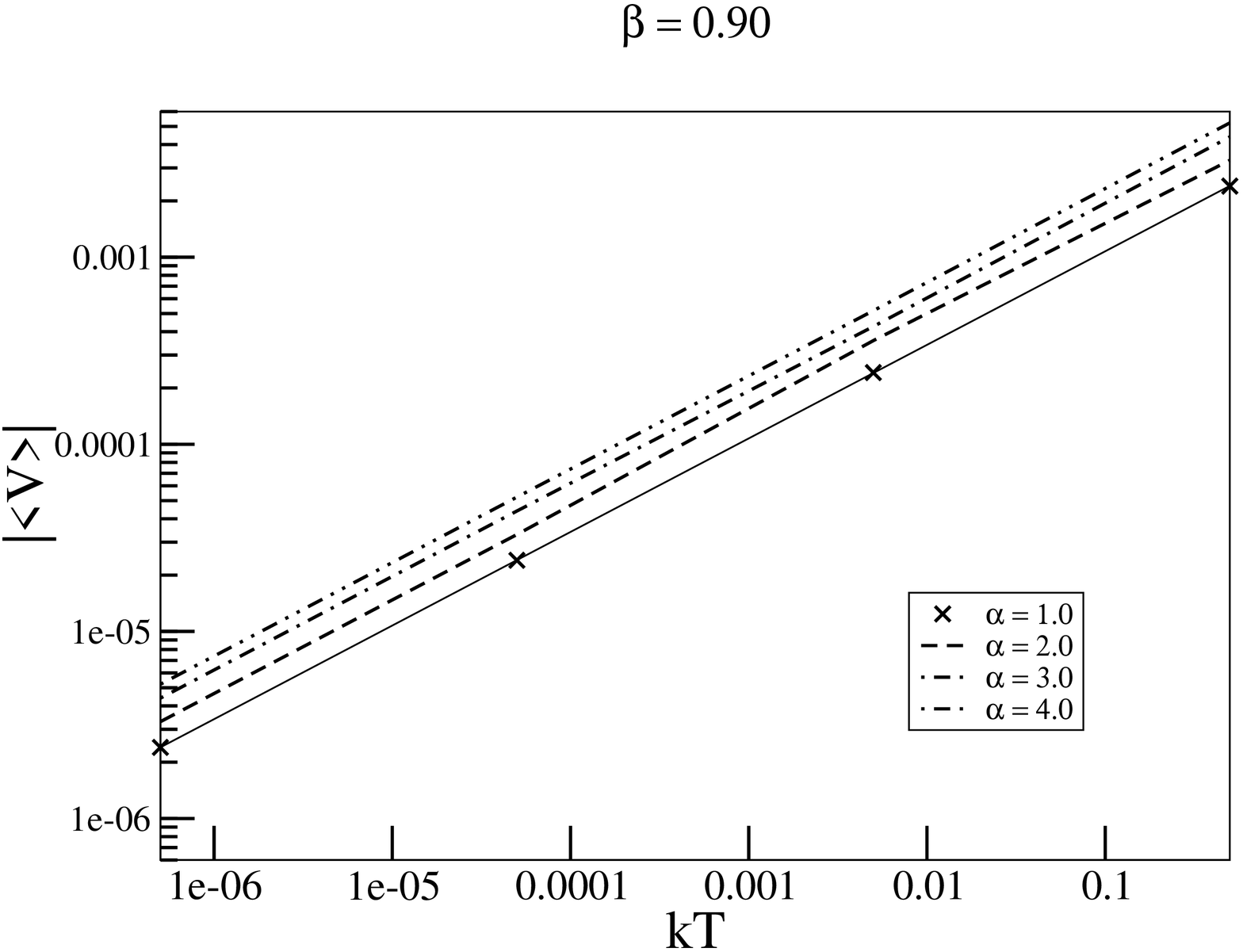}}
\caption [Figure 4]{ Scaling of $|<V>|$ with $k_BT$.}
\end{figure}
\par
Its important to note that, standard equilibrium statistical physics is actually in contradiction with Newton's first law if it totally denies the existence of any uniform motion under the force balanced situation. A mechanical equilibrium in general is a force balanced situation and some velocity gained during the initial transient to equilibrium should be maintained if under certain conditions equilibrium renders the force balance. Our Eq.6 is the statement of such a force balance where the damping of the velocity of the CM is equal to the average force acting on the system due to broken symmetry. Its kind of a self-consistency condition that selects the average velocity of the CM. Just because such a situation had previously not been conceived of, it cannot be enough reason to think that there cannot exist a situation under equilibrium which would be consistent with Newton's first law specially when the Hamiltonian dynamics is the basis of equilibrium statistics. The problem so far has been that the results of particle dynamics imposed to everything. We see that, the moment we start considering the structure within the particle, there are more scopes of having new results. Moreover, equilibrium being defined by a whole lot of stationary quantities, one should keep in mind that those stationarity are average stationarity. Thus, there is scope of having other structures withing the average time period over which this stationarity is defined. One such possibility has been discussed in the relation with inelastic collisions and possible relatively long lived fluctuations bringing in so far unknown results. In general, it appears interesting to look at the effect of these internal geometric and functional structures in the fluctuations within equilibrium, specially when, we are able to do experiments at this scale these days. It probably also would provide means of studying non-equilibrium and equilibrium on the same footing because when fluctuations are long-lived they actually enable us seeing non-equilibriums within equilibrium.
\\
{\bf Discussion}\\

Having understood the physics at the origin of sustained uniform motion in equilibrium let us emphasize on the fact that such a directed motion is perfectly consistent with the Kelvin statement of the second law of thermodynamics. Since, we are considering a closed system of the bath and the dimer, no energy is being extracted from the system and the situation in no way violates energy conservation. Any extraction of energy from such a system would start a transient where the bath would start losing temperature because no bath has infinite heat capacity and the system is no longer in equilibrium. Such a transient would persist up to the point where the velocity of the system (being proportional to the square root of the temperature) is not enough to drive it against a barrier and in the resulting equilibrium situation the dimer would come to rest. So, in no way there could be perpetual extraction of energy making the dimer move against a global barrier. Our model does not include such a situation, its an isolated bath dimer system not having any other external drives, barriers and dissipative interactions. 
\par
Let us also note that for a uniformly moving system of particles in equilibrium the Detailed Balance would also hold good due to the fact that homogeneity of space makes interactions between particles function of the separations between them. So, no contribution to the internal energy of the system is there due to CM motion. Equilibrium probability distribution, that ensures detailed balance, being a function of the interactions between particles would remain unaltered. This fact is explicit in the form of $P(Z)$ we have got. Moreover, the heat bath does not set any reference frame to the dimer means the probability of the CM to be found anywhere in the bath is a constant. This, in other words, is the homogeneity of space for the dimer. The homogeneity of space is also important to understand that no entropy production would be there for such a system under uniform motion in equilibrium. The rate of entropy ($S(t))$) production is given by the expression
\bea
\frac{\partial S(t)}{\partial t} = \int{dx(\ln{P(x,t)}+1)\frac{\partial P(x,t)}{\partial t}},
\eea
Where P(x,t) is the probability to have the configuration $x$ at time $t$. Now, the configuration being defined solely on the basis of the separation between particles and keeping homogeneity of space in mind (as our $P(Z)$), the $P(x,t)$ would never be an explicit function of time in equilibrium i.e. $P(x,t)=P(x)$. Thus, the partial time derivative of the configuration probability being zero the entropy is a constant which is the most important signature of equilibrium from which basically follows the detailed balance etc. Physically one can understand this constancy of the entropy from the constancy of the available phase space to the system. The conservation of the allowed momentum space to the system remains there in the presence of a constant drift of the CM. Actually the Maxwellian velocity distribution is centered around the CM velocity in such a situation and not around zero. The available volume to the system also remains constant since, due to uniform motion of the CM, as much new volume is getting available to the system as is lost at every infinitesimal time interval. This also keeps the available volume to the bath degrees of freedom constant.
\par
We first understand why the Smoluchowski-Feynman ratchet does not show directed transport in equilibrium whereas our system does. The reason of Smoluchowski-Feynman ratchet not showing directed transport lies in the fact that it is not at all a symmetry broken system when the temperatures of the two heat baths are the same i.e. in equilibrium. This fact has beautifully been demonstrated by Feynman showing that the probability of the pawl going up is the same irrespective of which bath it is driven by when the temperature of the baths are the same and that removes the broken symmetry \cite{feyn}. But, our system is a broken symmetric system perfectly consistent with the conditions of equilibrium and that is why it shows directed transport at constant temperature. So, directed transport in the absence of a driving, being in contact with a heat-bath, is possible. It does not automatically mean available energy and one should not confuse our present results with perpetual motion machines. Note that there exists example of uniform motion in equilibrium namely in superfluid helium and superconductors where a gap in the spectrum keeps the system away from excitations. So, one can reasonably understand that had the uniform motion and thermodynamic equilibrium been generically inconsistent because of detailed balance, entropy production etc., we would have not got such states. So, there exists no generic reason as to why one cannot expect such a situation in mechanical systems as well. Rather, our model shows how the detailed balance ensures broken symmetry, hence, directed motion keeping the energy balance intact.
\par
Le us have a look at very general dynamical system and its corresponding Fokker-Planck equation to have some clue as to how a directed motion is not at odds with the the demand of equilibrium. This part of the discussion would also help us understand what it actually means that the bath does not set any reference over space to the system. consider the most general form of dynamics of a particle as 
\ber\nonumber
\frac{\partial x(t)}{\partial t} &=& v(t)\\
\frac{\partial v(t)}{\partial t} &=& \Gamma v(t) -\frac{\partial V(x)}{\partial x} +f(t),
\eer
where in the above model, $V(x)$ is the potential accounting for all the forces other than the stochastic force $f(t)$ and the damping where $\Gamma$ is the damping constant. The Fokker-Planck equation for such a system can be immediately written (in 1D) as
\ber\nonumber
&&\frac{\partial P(x,v,t)}{\partial t} + v\frac{\partial P(x,v,t)}{\partial x}-\frac{1}{m}\frac{\partial V(x)}{\partial x}\frac{\partial P(x,v,t)}{\partial v} \\ &=& \frac{\Gamma}{m}\frac{\partial}{\partial v}\left [vP(x,v,t)+\frac{k_BT}{m}\frac{\partial P(x,v,t)}{\partial v}  \right ].
\eer
This Fokker-Planck equation has a solution of the form $P(x,v,t)=N\exp{(-V(x)/k_BT-mv^2/2k_BT)}$ where $N$ is the normalization constant. Had we done a Galilean transformation to our system Eq.10, its easy to see that we would have got a similar Fokker-Planck (due to $v$ appearing linearly in Eq.10) with $v$ replaced by $v+C$ (where C is the constant velocity) and the solution of the Fokker-Planck in that case would have been $P(x,v,t)=N\exp{(-V(x)/k_BT-m(v+C)^2/2k_BT)}$. So, the probability distribution has a uniform drift. Important to note that this uniform drift is actually through the bath although, of course, one would see the minimum of the global field $V(x)$ to move with the same velocity and the particle equilibrating at this minimum. But, this motion through the bath is something important to note which appears because of its homogeneity and of course not noticeable because of the same homogeneity. Nevertheless, this proves a point that under the above mentioned consideration of the dynamical presence of the heat bath with a damping and fluctuating force the rest of the system can actually move through it without experiencing any additional damping because of the baths homogeneity. This also indicates to the fact that such uniform motions can also be possibly present when the field is not fixed in space of the observer, but, can actually move with the system i.e. when the field $V(x)$ is due to internal interactions between a system of particles and has a minimum for the system to equilibrate in. Note that, the structure of the Fokker-Planck in such a case should remain the same with some additional terms on the right hand side due to the fact that internal field can now introduce some correlations between the velocities and a selection of the average velocity of the system should come from these additional terms. For our system with a hardcore collision i.e. not a smoothly varying confining field $V(x)$ it appears difficult to derive the related Fokker-Planck incorporating the velocity exchange at collision. Also note that, with inelastic collision and smoother collisions also its difficult to get the Fokker-Planck due to the extra dissipation present. However, We at present are doing work on it hopefully to get useful results which we can publish later. But, our present analysis being simple actually captures such a possible situation which is very much plausible in quite an easy way.
\par
The most important part in the construction of our model is the consideration of two time scales where at the slow scale the symmetry is broken and the fast scale is symmetric. This two time scale scenario can be considered as a generic mechanism for construction of such models. Important to note that the Langevin picture developed for a Brownian particle is based on integrating out the faster degrees of freedoms of the bath. But, in the case of two Brownian particles bound to each other where they frequently collide with each other and the collision time is much small compared to the Langevin time scale - that part of the fast dynamics has not been integrated out. It seems a simple but extremely effective technique to take into account the faster dynamics through the boundary conditions. The correctness of this technique ultimately has to be verified by experimental results. Note that, treating simultaneously two such widely different time scales and forces where the dynamics essentially becomes nonlinear is extremely difficult if not impossible. Appropriate boundary conditions based on experimental observations is one possible way out and here we are proposing one which definitely has to be checked by experiments. The source of our conviction at the present case is (a) the excellent agreements with the numerical results (b) collision does not feed energy to the system so that $<Z>$ can go on increasing for ever.
\par
In the present work we have argued that second law and detailed balance hold good. We have shown that the entropy remains constant as it should be in equilibrium. The only difference with the so far existing knowledge of equilibrium is that the time reversal symmetry present in stationary equilibrium is not there for uniformly moving systems. We know consequences of second law that fixes an arrow of time goes against time reversal symmetry. Its also important to note that the master equation for the probability distribution dynamics does not have time reversal symmetry. Then, why only a restricted class of solutions of the master equation with time reversal symmetry has to be considered as equilibrium solution particularly when uniformly moving ones are also entropy conserving? Moreover, a uniform motion is a mechanical equilibrium situation. It does not contribute to the temperature of the system because temperature is a measure of the width of velocity distribution. So, thermal equilibrium does not get perturbed by uniform motion of the whole system. There can indeed be stationary solutions of the master equation on uniformly moving frames which can always equally qualify as equilibrium solutions of the system having the time reversal asymmetry of the basic dynamics (master equation). An important point to be noted is that, within the scope of our present treatment, we never know what happens when the hydrodynamics of the heat bath becomes important because Langevin dynamics does not take that into account. Our analysis, as many others, is based on the assumption that in over-damped regime the effect of hydrodynamics is negligible. Also note that, under non-equilibrium conditions when we get directed transport (there are numerous papers on such driven diffusive systems) we take the driving to have no directional bias (that's what makes the result nontrivial) and still we get directed motion of the object under consideration in contact with the heat bath. So, a conservation of linear momentum would require an overall drift of the bath degrees of freedom in the direction opposite to that of the transporting system. If this effect is to be important to the extent it prevents the directed transport we have seen under equilibrium conditions, so it would do in the non-equilibrium conditions as well.
\par
To conclude, we would like to say that, with the help of a somewhat idealized model we probed the problem of having symmetry breaking directed motion in equilibrium because the model is exactly solvable. The physical understanding we gain from this analysis leads us to realize possible conditions under which such a motion can be physically realizable. We understand that dissipation plays a crucial role in such processes and based on which we can actually relax the stringent conditions of our idealized model. A possible schematic description of an experiment would be the following. If it is possible to implant a charge on one of the constituent particles of the dimer having the effective attractive interaction between the particles of the dimer by other means (e.g. effective hydrophobic interaction) then one can use a magnetic field to have such systems rotate on a circle while having some average velocity of the CM. Remember that the Lorenz force would not do any work on the system. Most probably one would have to give the system some initial velocity just to break the isotropy in the beginning and then the system should eventually settle to its average speed under the given conditions of the bath. We strongly feel that this is an area worth investigating because it can change the way we understand statistical physics at the present times.
\\
{\bf Acknowledgement}\\
I acknowledge having useful discussions with Prof. J.K. Bhattacharjee and Prof. Luca Peliti. I also acknowledge that an important issue raised by Prof. C.R. Doering helped rewriting this manuscript in a more clear way.

\newpage

\end{document}